**Title**

Interpolation-Split: a data-centric deep learning approach with big interpolated data to boost airway segmentation performance

Wing Keung Cheung[1,2], Ashkan Pakzad[1,3], Nesrin Mogulkoc[4], Sarah Needleman[1,3], Bojidar Rangelov[1,3], Eyjolfur Gudmundsson[1,2], An Zhao[1,2], Mariam Abbas[2], Davina McLaverty[5], Dimitrios Asimakopoulos[6], Robert Chapman[7], Recep Savas[8], Sam M Janes[9,10], Yipeng Hu[1,3], Daniel C. Alexander[1,2], John R Hurst[10,11], Joseph Jacob[1,9,10]

[1]Satsuma Lab, Centre for Medical Image Computing, University College London, London, UK

[2]Department of Computer Science, University College London, London, UK

[3]Department of Medical Physics and Biomedical Engineering, University College London, London, UK

[4]Department of Respiratory Medicine, Ege University Hospital, Izmir, Turkey

[5]Medical School, University College London, London, UK

[6]School of Clinical Medicine, University of Cambridge, Cambridge, UK

[7]Interstitial Lung Disease Service, Department of Respiratory Medicine, University College London Hospitals NHS Foundation Trust, London, UK

[8]Department of Radiology, Ege University Hospital, Izmir, Turkey

[9]Lungs for Living Research Centre, UCL, London, UK

[10]UCL Respiratory, University College London, London, UK

[11]Respiratory Medicine, Royal Free London NHS Foundation Trust, London, UK

Corresponding author:

Dr Joseph Jacob

UCL Centre for Medical Image Computing

1st Floor, 90 High Holborn, London WC1V6LJ

j.jacob@ucl.ac.uk

**Abstract**

The morphology and distribution of airway tree abnormalities enables diagnosis and disease characterisation across a variety of chronic respiratory conditions. In this regard, airway segmentation plays a critical role in the production of the outline of the entire airway tree to enable estimation of disease extent and severity. In this study, we propose a data-centric deep learning technique with big interpolated data, Interpolation-Split, to boost segmentation performance of the airway tree. The proposed technique utilises interpolation and image split to improve data usefulness and quality. Then, an ensemble learning strategy is implemented to aggregate the segmented airway segments at different scales. In terms of average segmentation performance (dice similarity coefficient, DSC), our method (A) achieves 90.55%, 89.52% and 85.80%, (B) outperforms the baseline models by 2.89%, 3.86% and 3.87% on average and (C) produces maximum segmentation performance gain by 14.11%, 9.28% and 12.70% for individual case when (1) nnU-Net with instant normalisation and leaky ReLU, (2) nnU-Net with batch normalisation and ReLU and (3) modified dilated U-Net are used respectively. Furthermore, our proposed technique has a low RAM/GPU memory usage; and it is GPU memory efficient and highly flexible enabling it to be deployed on any 2D deep learning model.

**Introduction**

Abnormal dilatation of the airways is a key feature in the diagnosis of idiopathic pulmonary fibrosis (IPF) patients. Disease extent and severity in IPF can be assessed by visual analysis of high-resolution CT images by radiologists. This approach, however, is subjective and time consuming. Automated airway tree analysis [1, 2] is an alternative method which enables an objective quantitative assessment of airway damage and therefore disease severity in IPF. The key component of airway tree analysis is establishing the 3D geometry of the airway tree and the standard approach to obtain the airway tree is image segmentation.

Airway segmentation is an active research area [3]. The goal is to produce a complete airway tree including the trachea, bronchi, bronchioles and terminal bronchioles. The segmentation task is challenging as the intensity, scale/size and shape of airway segments and their walls change across generations. Classical segmentation methods such as the Frangi filter [4, 5] and region growing method [6] were first used to segment the airway tree. The Frangi enhancement filter constructs a Hessian matrix to extract tubular-like tissues (i.e., airways) and remove non-tubular tissues (i.e. lung). This approach shows promise on airway segmentation. However, the segmented airway tree is limited to the first few branching airway generations (i.e. between $1^{st}$ and $6^{th}$ generations). Furthermore, it requires tuning the parameters ($\alpha$, $\beta$ and $\sigma$) manually for extracting the optimal airway tree. This process is time-consuming and not user-friendly for clinicians. Employing a region growing algorithm is another approach to segment the airway tree. A seed point is first placed at the trachea, then the region is grown by adding neighbour voxels with a predefined intensity. The algorithm stops when no more voxels can be added. There are several drawbacks to this approach. Intensity thresholding is used to select voxels, but causes leakage (over-segmentation) when an aggressive threshold is used. Conversely, the airway is under segmented when a conservative threshold is used. Therefore, the completeness of the airway tree produced by this approach is limited.

Recent advances in deep learning provides new opportunities for segmentation. It utilises data and GPU technology and offers a fast and fully automatic method to perform segmentation. Deep learning can be divided into two branches – (1) Model-centric and (2) Data-centric. Model-centric deep learning focuses on the model architecture and keeps the data unchanged. Popular models have been developed to tackle segmentation challenge. For examples, SegNet [7] and HRNet [8] are proposed for general segmentation. Unet [9] and Vnet [10] are deployed for medical image segmentation. These models produce a good segmentation though they require high GPU memory usage. On the other hand, data-centric deep learning focuses on the data and keeps the model unchanged. Data augmentation [11] is an example of manipulating the source data to produce more varied samples. It uses geometrical transformation (i.e., flip, rotate and crop) to modify the images. The model performance can be improved by training on a dataset with richer features. Active learning [12] is another example of a data-centric technique. It aims to select the most useful data for labelling and permits the user to interact with the deep learning model to complete the data annotation. This technique improves the efficiency of the annotation task. Furthermore, a data-centric deep learning approach is particularly attractive as it requires low GPU memory usage and its approach is straightforward to implement.

The studies related to model-centric deep learning in airway segmentation are summarised below. A convolutional neural network (CNN) based leak detection method to improve airway segmentation was proposed by Charbonnier et al. [13]. Yun et al. [14] presented a 2.5D CNN for airway segmentation. This approach achieved about 90% DSC accuracy. A 3D Unet to detect topological leak was employed by Nadeem et al. [15]. The intensity threshold was adjusted on the probability map and a freeze-and-growth algorithm was used to correct the leaks. Qin et al. [16] developed a simple-yet-effective deep learning method for this task. It utilised a context-scale fusion strategy to improve the connectivity between airway segments. The DSC of this approach is 93% on a public dataset. A three-dimensional multi-scale feature aggregation network was proposed by Zhou et al. [17] to handle the difference in scale of substructures during airway tree segmentation. This method produced results with 86.18% DSC and 79.31% true positive rate (TPR). Further, a simple and low-memory 3D Unet was developed by Garcia-Uceda et al. [18]. It processed large 3D image patches in a single pass within the network creating a robust and efficient analysis. Zheng et al. [19] proposed WingsNet with group supervision to deal with class imbalance between airway and non-airway regions. The branch detection rate of the proposed method is 80.5%. A coarse-to-fine segmentation framework was deployed by Guo et al. [20]. This utilised a multi-information fusion convolution neural network (Mif-CNN) and a CNN-based region growing for main airway and small branch segmentation. The DSCs of this work were 93.5% and 95.8% for private and public datasets respectively. Wang et al. [21] developed a spatial fully connected tubular network with a novel radial distance loss for 3D tubular-structure segmentation. The method provided a better airway tree segmentation than the baseline Unet model. A joint 3D UNet-Graph Neural Network-based method was presented by Juarez et al. [22]. It used graph convolutions to improve airway connectivity. More recently, Zheng et al. [19] identified the gradient erosion and dilation problem and designed a group supervision to enhance the training of the network. A general union loss was also developed to tackle the intra-class imbalance issue through distance-based weights and element-wise focus on the hard-to-segment regions. Wu et al. [23] proposed a long-term slice propagation method for airway segmentation. The method achieved 92.95% DSC. A novel label refinement method was developed by Chen et al. [24] to correct the structural errors in airway segmentation. It produced an airway segmentation with DSC between 79% and 81%. Zhao et al. [25] developed Group Deep Dense Supervision for small bronchiole segmentation. This method has a high sensitivity in detecting fine-scale branches and outperforms state-of-the-art methods by a large margin (+12.8% in Branch Detection and +8.8% in Tree Detection).

Interpolation has been widely used in image processing. The mechanism of interpolation involves resampling, several interpolating functions have been used for image resampling [26], i.e., nearest neighbour function, linear function and cubic B-spline function. Interpolation has also been used in image augmentation [27]. It is applied to either input space or feature space. The purpose of this technique is to diversify the training samples by manipulating features in the input or feature spaces and hence improve the generalisation. Furthermore, it can be used to filling-in the blank part of the image after image manipulation [28]. i.e., rotation. Cropping is used in conjunction with interpolation to achieve the desirable results. Existing techniques such as random scaling, random cropping and random cropping with scaling can increase the variability of the training images. For example, RandomResizedCrop from PyTorch. It crops the image randomly and the sub-image is subsequently up-scaled to the original image size by interpolation. The drawback of this approach is that the random cropping can miss the important features in the image and the up-scaling can increase the blurring and edge effects on the sub-image. To resolve these issues, a novel technique, Interpolation-Split, is proposed in this study. It performs systematic up-scaling and followed by systematic splitting on the image. In the context of airway segmentation, this new approach can ensure all airways are captured and utilised. Further, it minimises blurring and edge effects when interpolation is performed.

Additionally, no study focuses on a purely data-centric approach for airway segmentation. Therefore, in this study, we propose a 2D data-centric deep learning method for the automated segmentation of airway trees on HRCT images. The proposed technique is evaluated by comparing the segmentation performance with three baseline models (2D nnU-Net with instant normalisation (IN) and leaky ReLU, 2D nnU-Net with batch normalisation (BN) and ReLU and 2D modified dilated U-Net).

The main contributions of this study are:

- The first study to propose a 2D data-centric deep learning method with interpolation that segments the airways on HRCT images.

- The proposed technique utilises interpolation and image split to improve data usefulness and quality.

- The study combines big interpolated data (972655 samples) and data-centric deep learning method to boost airway segmentation performance.

- An ensemble learning strategy is implemented to aggregate the segmented airway segments at different scales.

- The proposed technique has a low RAM/GPU memory usage, it is GPU memory efficient and highly flexible to be deployed in any 2D deep learning model.

**Methods**

*Clinical data*

The clinical data (n = 30) contained healthy subjects, patients with heart disease and patients with IPF. It included a healthy subject and 6 patients with heart disease or IPF from the EXACT09 dataset [29], 6 healthy never-smoker subjects and 17 IPF patients from University College London Hospital. The study was carried out in accordance with the recommendations of University College London Research Ethics Committee, with written informed consent from all subjects. The data including source images and their ground-truth masks were further divided into training (66%) and validation (34%) sets. Table 1 shows the subject/patient information in the validation set. The number of samples (source images) for training and validation is shown in Table 2.

Table 1: The subject/patient information in the validation set.

| Subject/Patient | Status |
|---|---|
| **case 1** | Healthy |
| **case 2** | Healthy |
| **case 3** | Patient with IPF |
| **case 4** | Patient with heart disease |
| **case 5** | Patient with IPF |
| **case 6** | Patient with IPF |
| **case 7** | Healthy |
| **case 8** | Patient with IPF |
| **case 9** | Patient with IPF |
| **case 10** | Patient with heart disease |

Table 2: The number of samples (source images) for training and validation.

| **Interpolation ratio ($ir$)** | **Training set** | **Validation set** |
|---|---|---|
| 1 (original dataset) | 7552 | 3891 |
| 2 | 30208 | 15564 |
| 4 | 120832 | 62256 |
| 8 | 483328 | 249024 |
| **Total** | 641920 | 330735 |

*Data pre-processing*

The data were pre-processed in three steps: (1) ImageJ was used to convert the source images from DICOM format to TIFF format. (2) The images were subsequently normalised by using the following settings to emphasize lung tissue visualisation: W = 1500 HU, L = -500 HU. (3) The intensity of the normalised images were rescaled in the range 0 to 255HU. The annotation of the ground-truth mask was performed on 3D Slicer.

*Interpolation-split*

The details of Interpolation-Split is as follows. The CT image and its mask are zoomed in at various scales. The zoomed-in CT images and masks are produced by interpolation and split. The original CT images are up-sampled by bi-linear interpolation, while the original masks are up-sampled by nearest neighbour interpolation. Then, the interpolated image is split into sub-images with fixed dimensions (512x512). Here, an interpolation ratio (*ir*) is defined to control the zoom-in scale. For example, the dimension of the interpolated image (1024x1024) is doubled from the original image (512x512) when *ir* is set to 2. Then, the interpolated image is split into four sub-images (512x512). The interpolation and split mechanism (i.e., ir2) is demonstrated in Figure 1. Further, the effect of the interpolated ratio (*ir* = 2, 4 and 8) is investigated. It should be noted that no interpolation and split is performed for *ir* = 1.

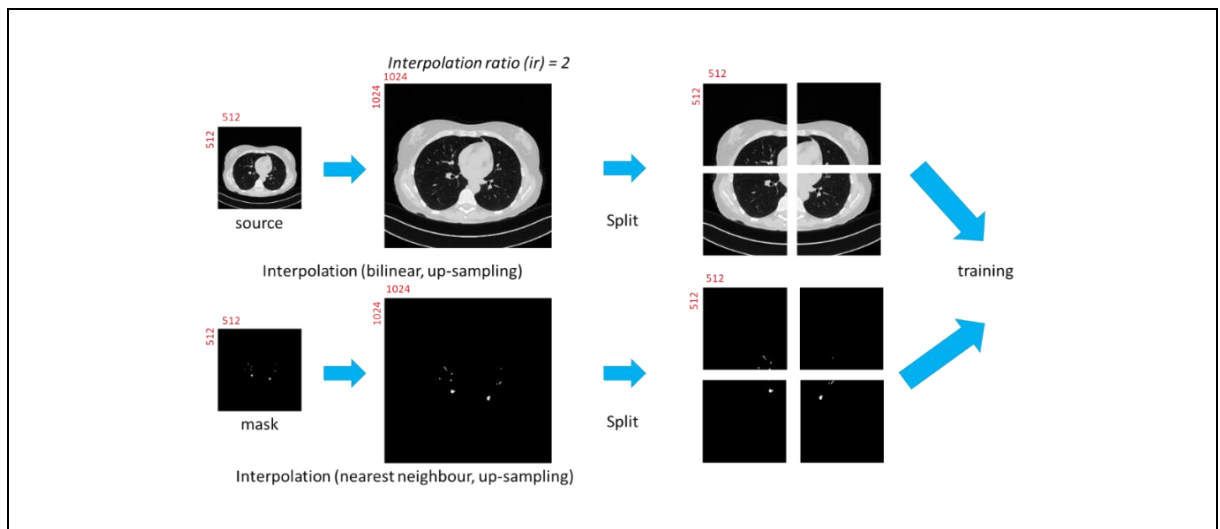

Figure 1: The interpolation and split mechanism.

*Blurring and edge effects*

The blurring and edge effects were compared between the existing technique and Interpolation-Split. A slice was selected from each case, then a set of sub-images were created by the existing technique and Interpolation-Split. (1) Existing technique: a single image (512x512) was cropped into 64 sub-images (8x8), then each sub-image was up-scaled to the original size (512x512). (2) Interpolation-Split: a single image (512x512) was up-scaled to 4096x4096 (ir8), then the interpolated image was split into 64 sub-images (512x512). Four sub-images from each case were selected for comparison. In total, 120-paired sub-images were produced. Diagonal Laplacian [30] was employed to measure the sharpness of the sub-image. Further, a paired t-test was used to evaluate whether there is any statistical significance in sharpness between the sub-images produced by the existing technique and Interpolation-Split. A *p*-value < 0.05 was considered significant for statistical analysis. The analysis was performed on SPSS (version 27, IBM).

*Selected models for performance evaluation*

Three state-of-the-art models were selected for evaluating our proposed method. These 2D models are (A) nnU-Net with instant normalisation and leaky ReLU (B) nnU-Net with batch normalisation and ReLU (C) modified dilated U-Net.

(1) nnU-Net
nnU-Net [31] is a deep learning based semantic segmentation method. It offers automatic configuration including pre-processing, network architecture, training and post-processing for any segmentation task. In this study, two network configurations, instant normalisation with leaky ReLU and batch normalisation with ReLU, were chosen to evaluate our Interpolation-Split method.

(2) modified dilated U-Net
The airway was segmented by using a modified dilated U-Net. A dilated U-Net is an extended model of the original U-Net [9] and adopts an encoder-decoder architecture. The encoding path captures features from images and the decoding path localises these features. A sequential dilation module [32] is employed in the bottleneck layer and this improves global context capturing and maintains the resolution of the feature map. Furthermore, the dilated U-Net was modified by introducing batch normalisation and dropout. These modifications improve model stability and segmentation performance. The schematic diagram of the modified dilated U-Net and the sequential dilation module are shown in Figures 2 and 3.

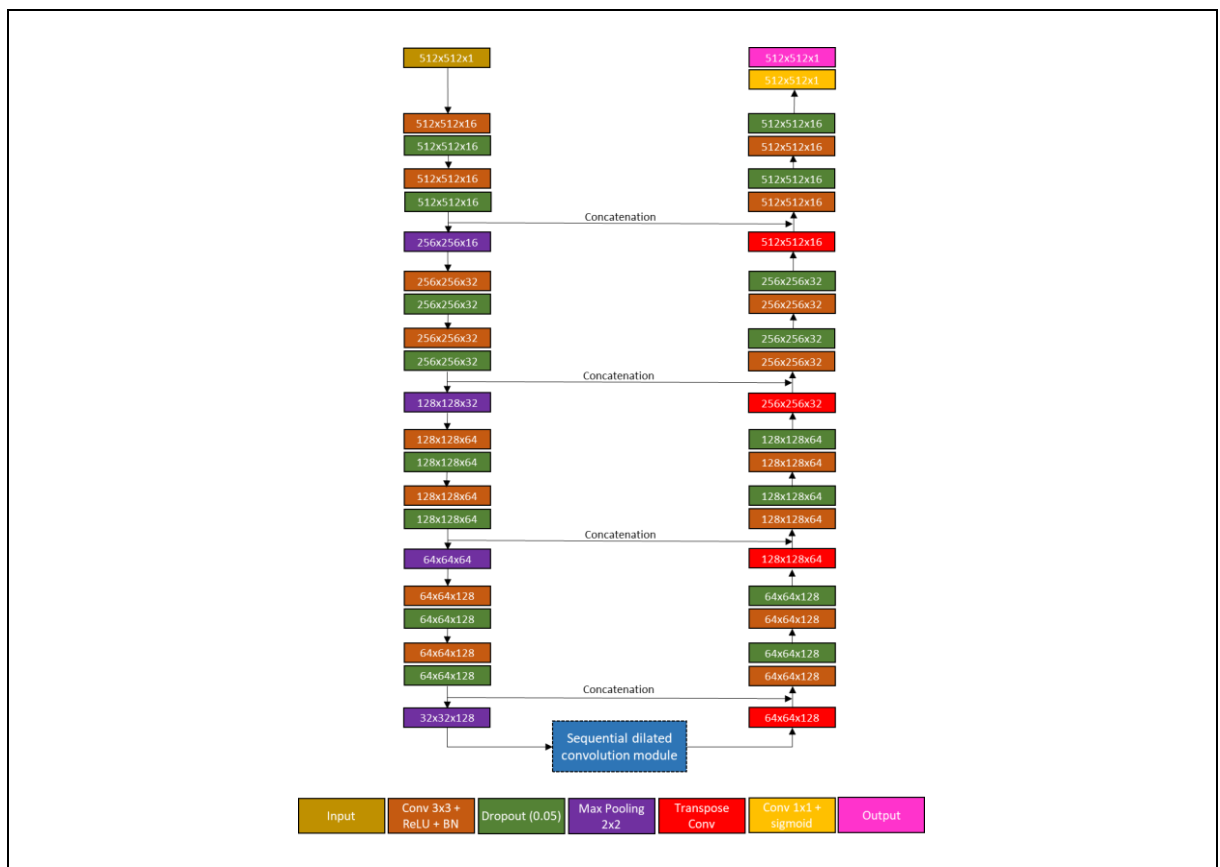

Figure 2: The network architecture of the modified dilated U-Net.

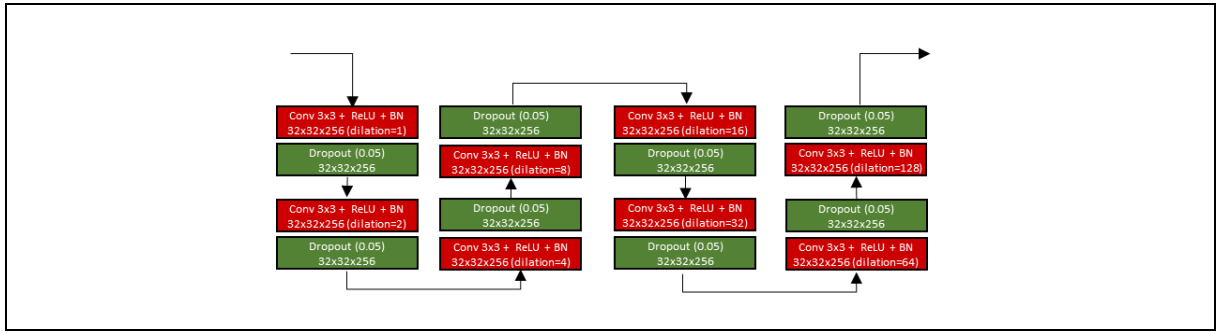

Figure 3: The sequential dilated convolution module.

*Implementation and training strategy*

The proposed models (per *ir*) were trained and implemented on a high performance cluster with deep learning frameworks installed. Specifically, PyTorch (v2.0.1), Tensorflow (v1.1.4) and Keras (v2.2.4) were executed on Linux (Rocks 7.0/CentOS 7.9.2009). Furthermore, various computing machines with Intel/AMD multi-core CPU chipset and Nvidia GPU cards were used to complete the training.

nnU-Net provided an automatic configuration for models training. The configuration includes fixed, rule-based and empirical parameters. The setting of fixed parameters is shown in Table 3.

Table 3: The setting of fixed parameters for models (nnU-Net) training.

| **SGD with Nesterov momentum ($\mu = 0.99$) optimiser** | |
|---|---|
| Learning rate (Poly learning rate schedule, initial) | 0.01 |
| Epochs | 1,000 epochs $\times$ 250 mini-batches |

The trained models for modified Dilated U-Net were produced by employing the Adam optimiser, ReduceLROnPlateau and Early stopping. The setting of parameters is shown in Table 4.

Table 4: The setting of parameters for models (modified Dilated U-Net) training.

| **Adam optimiser** | |
|---|---|
| Learning rate (initial) | $10^{-3}$ |
| Epochs | 200 |
| | |
| **ReduceLROnPlateau** | |
| Factor | $10^{-1}$ |
| Patience | 3 |
| Min_lr | $10^{-5}$ |
| | |
| **Early stopping** | |
| Patience | 10 |

The loss function, combined loss, was used to train the deep learning model. The combined loss function includes binary cross entropy (BCE) and dice similarity coefficient (DSC). The BCE is used to calculate the difference between the two probability distributions (foreground vs background) while DSC is used to measure the similarity between predicted segmentation and the ground-truth segmentation.

The prediction (per *ir*) was done using the trained models above. The unseen source images were interpolated and split to form the inputs for model prediction. When the prediction was complete, the initial predicted masks were merged and down-sampled (nearest neighbour) to the final mask with size 512x512. The workflow of the prediction mechanism (i.e., ir2) is shown in Figure 4.

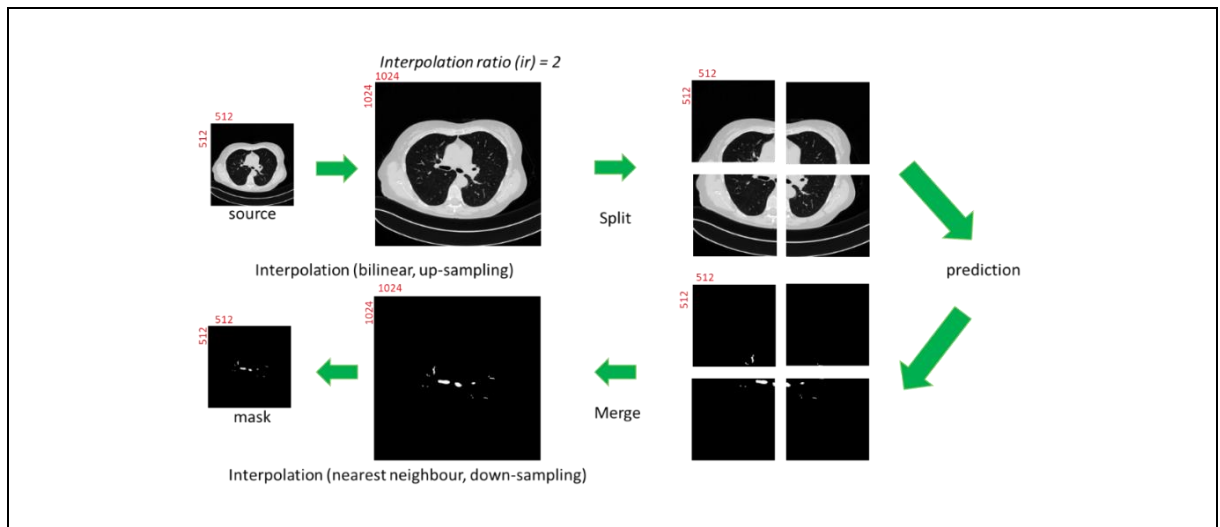

Figure 4: The workflow of prediction mechanism.

*Ensemble learning strategy*

The baseline model (ir1) has the ability to segment the airway from trachea to 6-8 airway generations, while those 9 or above airway generations are missed. An ensemble learning strategy is proposed to overcome the segmentation limitation. By increasing *ir* (i.e., ir2, ir4 and ir8), the optimal segmented airway is shifted towards the airway with smaller diameter or higher generation. Then, the optimal segmented airway with various *ir* are aggregated. Finally, the airway tree with higher generations (9 or above) is produced.

The segmented masks from *ir* = 1, 2, 4 and 8 are aggregated to form a combined mask. This is done by applying a union operation on all mask sets. Finally, the largest connected component of an airway in the combined mask is extracted and hence the final segmented mask is produced. The workflow of this ensemble learning strategy (i.e., ir1 + ir2 + ir4 + ir8) is shown in Figure 5.

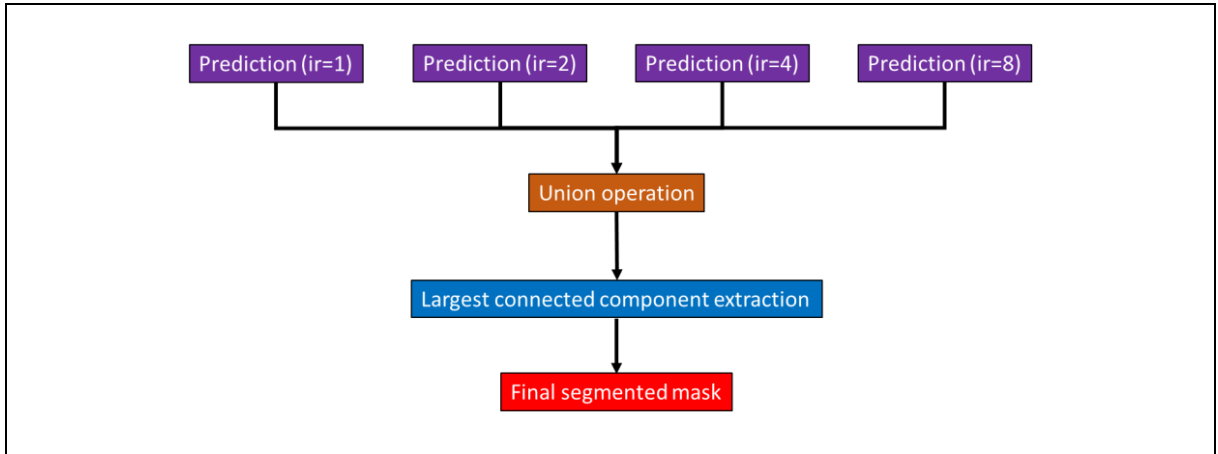

Figure 5: The workflow of ensemble learning strategy.

## Results

*Blurring and edge effects*

The mean sharpness of sub-images (n = 120) was 1.62 ± 0.43 produced by Interpolation-Spilt and 1.59 ± 0.42 produced by existing technique ($p < 0.001$). Our proposed technique produced less blurry images than the existing technique. An example of the edge effect was demonstrated in Figure 6. Our Interpolation-Split produced a better sub-image with minimal edge effect.

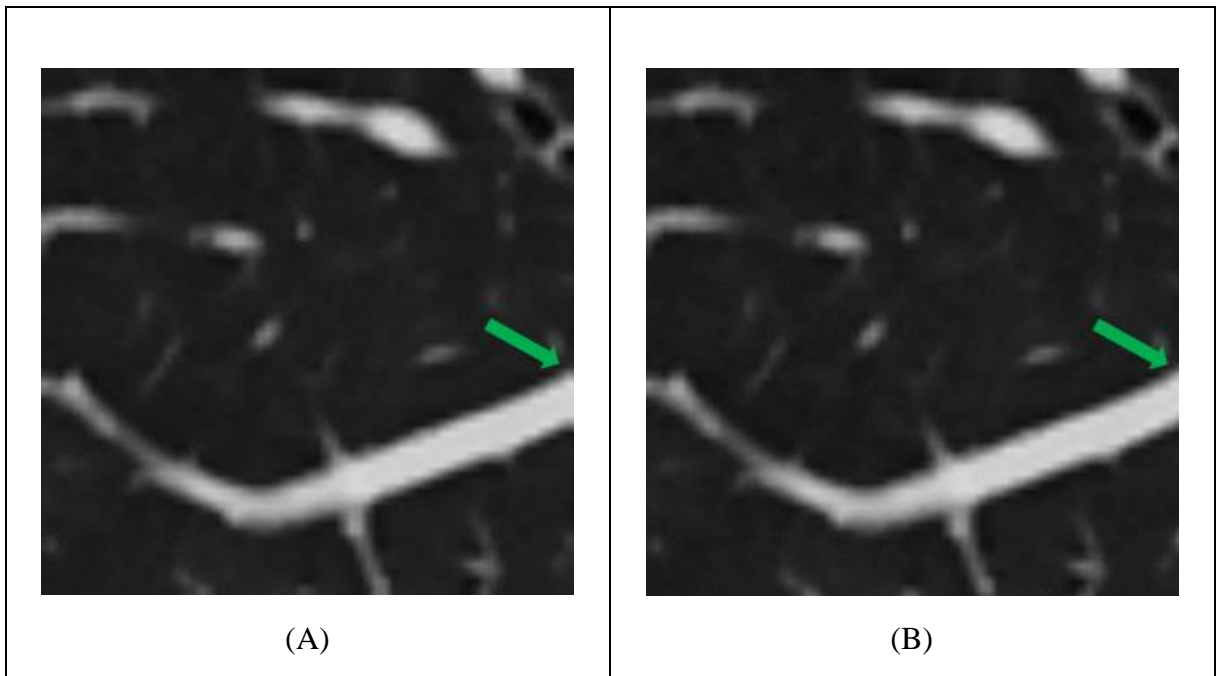

Figure 6: Edge effect (A) The sub-image produced by existing technique and a missing tissue (green arrow) was seen on the edge, (B) The sub-image produced by Interpolation-Split and a smoother tissue boundary (green arrow) was seen.

*Airway segmentation performance*

Tables 5 shows the airway segmentation performance by using state-of-the-art models - nnU-Net with IN and leaky ReLU, nnU-Net with BN and ReLU and modified dilated U-Net . Our proposed data-centric method provides a better airway segmentation compared to a baseline model (ir1) for all models. On average, our Interpolation-Split (ir1 + ir2 + ir4 + ir8) with nnU-Net with IN and leaky ReLU has the highest DSC (90.55%), while the DSC of nnU-Net with BN and ReLU and modified dilated U-Net are 89.52% and 85.80% respectively.

Table 5: Airway segmentation performance in percentage.

| DSC | nnU-Net (IN + leaky ReLU) | | nnU-Net (BN + ReLU) | | modified dilated U-Net | |
|---|---|---|---|---|---|---|
| | baseline | ir1 + ir2 + ir4 + ir8 | baseline | ir1 + ir2 + ir4 + ir8 | baseline | ir1 + ir2 + ir4 + ir8 |
| case 1 | 87.06 | 86.85 | 86.65 | 86.86 | 85.27 | 86.10 |
| case 2 | 82.33 | 83.14 | 81.89 | 83.29 | 81.56 | 81.81 |
| case 3 | 77.96 | 81.23 | 78.01 | 81.58 | 78.99 | 81.40 |
| case 4 | 86.50 | 88.29 | 87.24 | 88.60 | 83.20 | 86.75 |
| case 5 | 88.07 | 88.17 | 87.23 | 87.50 | 79.71 | 85.21 |
| case 6 | 96.58 | 98.06 | 93.70 | 95.48 | 88.69 | 93.28 |
| case 7 | 92.38 | 94.09 | 86.39 | 92.40 | 78.56 | 79.80 |
| case 8 | 97.00 | 97.35 | 88.31 | 97.19 | 91.13 | 94.55 |
| case 9 | 83.47 | 97.58 | 82.04 | 91.32 | 72.59 | 76.80 |
| case 10 | 85.22 | 90.74 | 85.12 | 90.93 | 79.57 | 92.27 |
| Average ± SD | 87.66 ± 6.12 | 90.55 ± 6.06 | 85.66 ± 4.27 | 89.52 ± 4.96 | 81.92 ± 5.39 | 85.80 ± 6.04 |

The airway segmentation results of case 6 and case 9 are shown in Figures 7 and 8. For the DSC of case 6, our method achieves 98.06%, 95.48% and 93.28% for nnU-Net (IN + leaky ReLU), nnU-Net (BN + ReLU) and modified dilated U-Net respectively. Regarding the DSC of case 9, our method achieves 97.58%, 91.32% and 76.80% for nnU-Net (IN + leaky ReLU), nnU-Net (BN + ReLU) and modified dilated U-Net respectively. Visually, the trachea and bronchi are well segmented for both cases. The majority of bronchioles are better segmented by our method.

| Model | Ground-truth | Baseline | Interpolation-Split (ir1 + ir2 + ir4 + ir8) |
|---|---|---|---|
| **nnU-Net (IN + leaky ReLU)** | 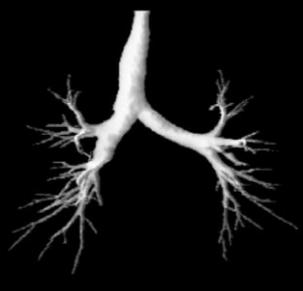 | 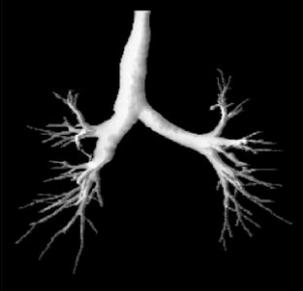 | 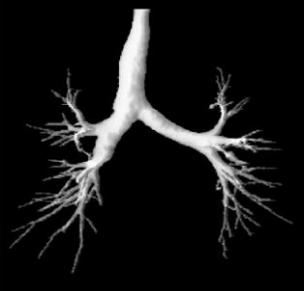 |
| **nnU-Net (BN + ReLU)** | 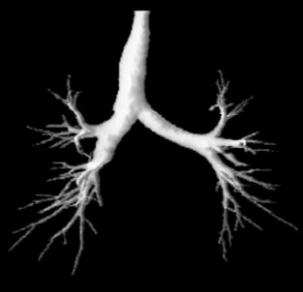 | 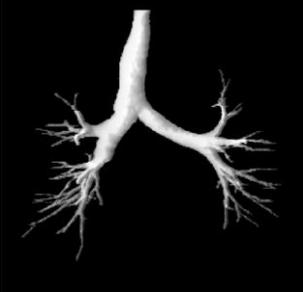 | 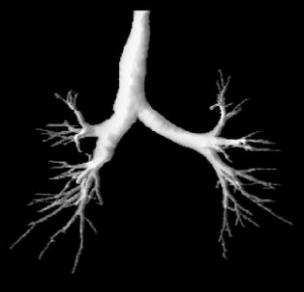 |
| **modified dilated U-Net** | 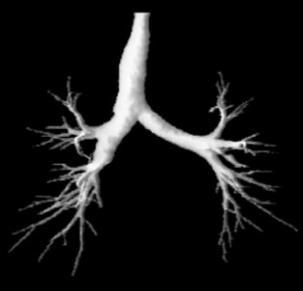 | 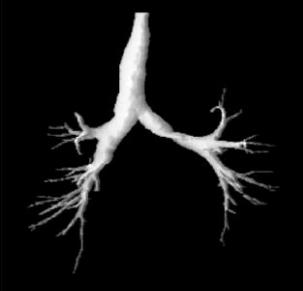 | 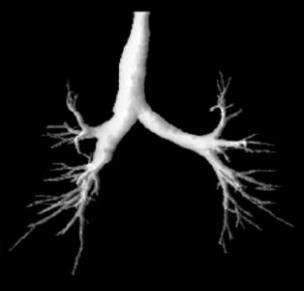 |

Figure 7: The airway segmentation of case 6 by our method for nnU-Net (IN + leaky ReLU), nnU-Net (BN + ReLU) and modified dilated U-Net.

| Model | Ground-truth | Baseline | Interpolation-Split (ir1 + ir2 + ir4 + ir8) |
|---|---|---|---|
| nnU-Net (IN + leaky ReLU) | 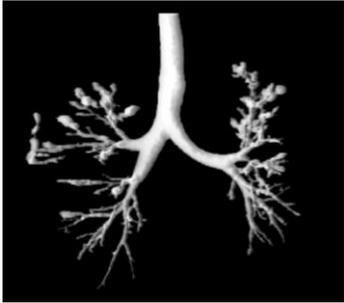 | 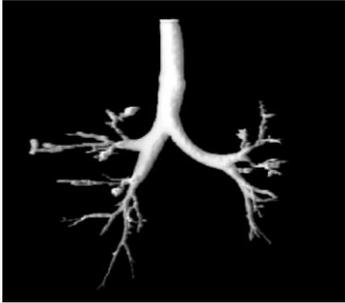 | 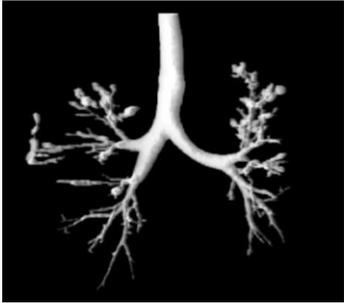 |
| nnU-Net (BN + ReLU) | 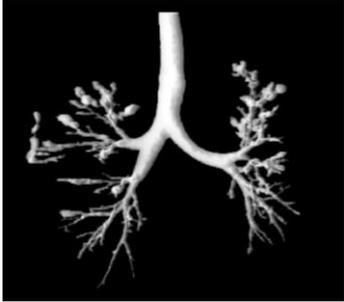 | 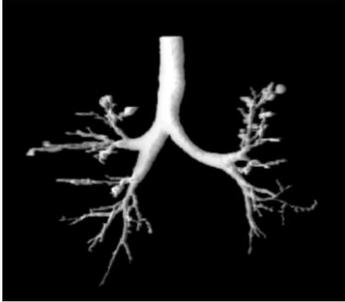 | 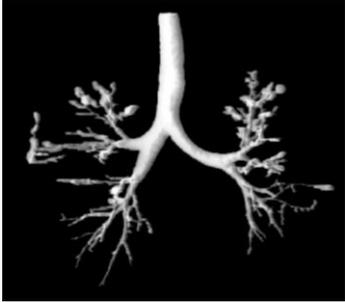 |
| modified dilated U-Net | 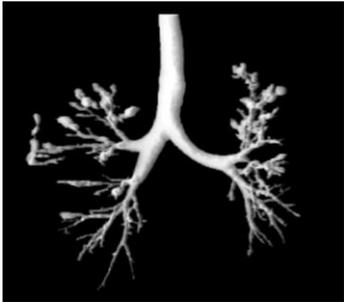 | 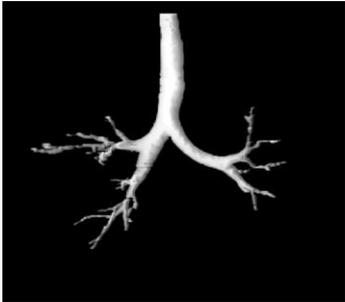 | 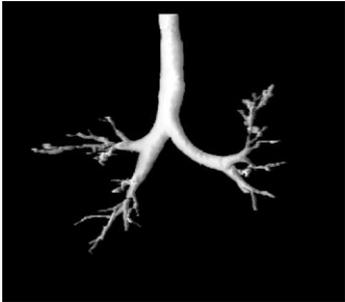 |

Figure 8: The airway segmentation of case 9 by our method for nnU-Net (IN + leaky ReLU), nnU-Net (BN + ReLU) and modified dilated U-Net.

*Airway segmentation performance gain*

The airway segmentation performance gain (expressed as a percentage) by using our method is reported in Table 6. On average, our Interpolation-Split (ir1 + ir2 + ir4 + ir8) with modified dilated U-Net has the highest average performance gain (3.87%), while the average performance gain of nnU-Net with BN and ReLU and nnU-Net with IN and leaky ReLU are 3.86% and 2.89% respectively. Notably, for highest segmentation performance gain of individual case, our method achieves 14.11% (case 9), 9.28% (case 9) and 12.70% (case 10) for nnU-Net (IN + leaky ReLU), nnU-Net (BN + ReLU) and modified dilated U-Net respectively.

Table 6: Airway segmentation performance gain/loss in percentage.

| performance gain/loss | nnU-Net (IN + leaky ReLU) baseline vs ir1 + ir2 + ir4 + ir8 | nnU-Net (BN + ReLU) baseline vs ir1 + ir2 + ir4 + ir8 | modified dilated U-Net baseline vs ir1 + ir2 + ir4 + ir8 |
|---|---|---|---|
| case 1 | -0.22 | 0.21 | 0.84 |
| case 2 | 0.82 | 1.40 | 0.25 |
| case 3 | 3.27 | 3.56 | 2.41 |
| case 4 | 1.79 | 1.36 | 3.56 |
| case 5 | 0.97 | 0.27 | 5.50 |
| case 6 | 1.48 | 1.78 | 4.59 |
| case 7 | 1.70 | 6.01 | 1.24 |
| case 8 | 0.35 | 8.87 | 3.42 |
| case 9 | 14.11 | 9.28 | 4.21 |
| case 10 | 5.51 | 5.81 | 12.70 |
| Average ± SD | 2.89 ± 4.29 | 3.86 ± 3.43 | 3.87 ± 3.54 |

Figure 9 shows the comparison of airway segmentation between our method (ir1 + ir2 + ir4 + ir8) and the baseline model (ir1) for cases 6 and 9. It is clear that our method segments more bronchioles than the baseline model. Furthermore, our method improves the airway wall segmentation for case 9.

| Model | Case 6 | Case 9 |
|---|---|---|
| **nnU-Net (IN + leaky ReLU)** | 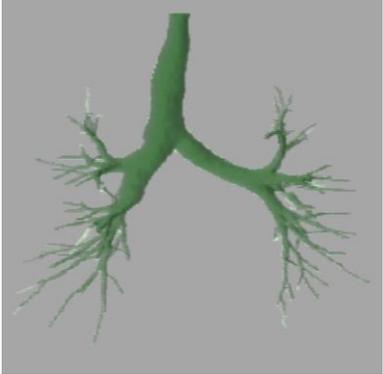 | 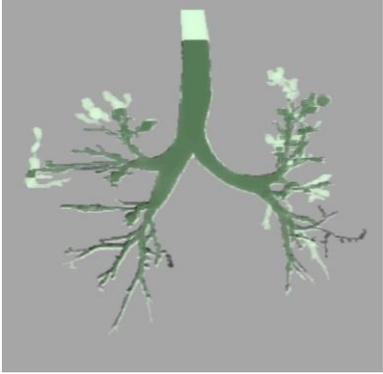 |
| **nnU-Net (BN + ReLU)** | 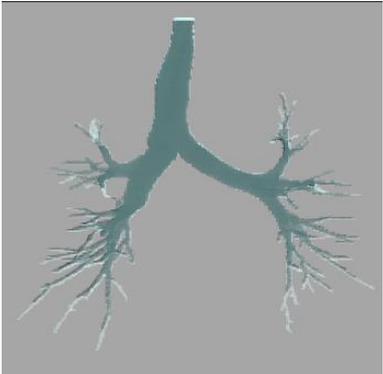 | 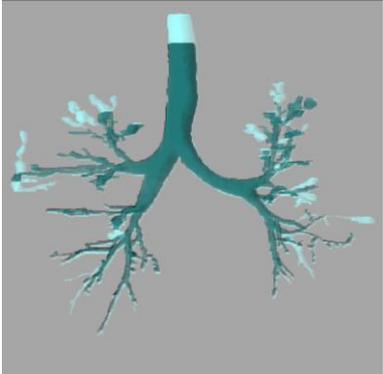 |
| **modified dilated U-Net** | 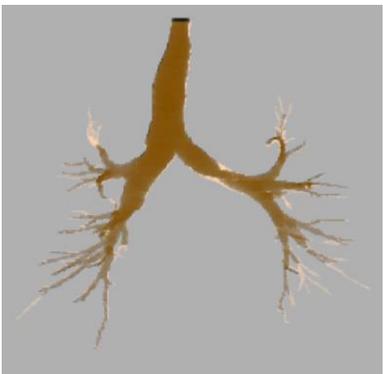 | 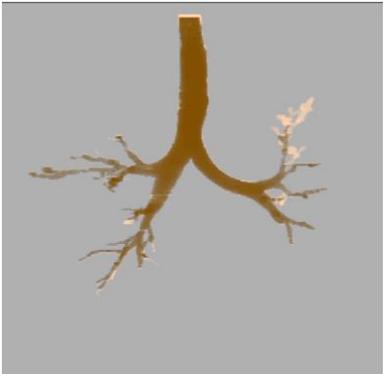 |

Figure 9: The comparison of airway segmentation between our Interpolation-Split (ir1 + ir2 + ir4 + ir8 - light green / light blue / light brown) and the baseline models (ir1 – green / blue / brown) for cases 6 and 9.

*The ablation study of the proposed method*

Table 7 shows the ablation study of the proposed method with four interpolation ratios applied to nnU-Net (IN + leaky ReLU), nnU-Net (BN + ReLU) and modified dilated U-Net. The average segmentation performance gain is improved when segmentation with a higher interpolation ratio is aggregated for all models. Further, these results confirm that our ensemble learning strategy works well.

Table 7: Ablation study of the proposed method with four interpolation ratios applied to nnU-Net (IN + leaky ReLU), nnU-Net (BN + ReLU) and modified dilated U-Net.

| Model | Ablation | | | | DSC |
|---|---|---|---|---|---|
| | ir1 | ir2 | ir4 | ir8 | (Average ± SD) |
| nnU-Net (IN + leaky ReLU) | ✓ | ✓ | ✗ | ✗ | 89.95 ± 5.90 |
| nnU-Net (IN + leaky ReLU) | ✓ | ✓ | ✓ | ✗ | 90.48 ± 6.01 |
| nnU-Net (IN + leaky ReLU) | ✓ | ✓ | ✓ | ✓ | 90.55 ± 6.06 |
| nnU-Net (BN + ReLU) | ✓ | ✓ | ✗ | ✗ | 89.19 ± 4.95 |
| nnU-Net (BN + ReLU) | ✓ | ✓ | ✓ | ✗ | 89.52 ± 4.94 |
| nnU-Net (BN + ReLU) | ✓ | ✓ | ✓ | ✓ | 89.52 ± 4.96 |
| modified dilated U-Net | ✓ | ✓ | ✗ | ✗ | 84.30 ± 5.96 |
| modified dilated U-Net | ✓ | ✓ | ✓ | ✗ | 85.27 ± 6.42 |
| modified dilated U-Net | ✓ | ✓ | ✓ | ✓ | 85.78 ± 6.04 |

*Effect of aggregated interpolation ratio (ir)*

The plot of average performance gain versus aggregated interpolation ratio is shown in Figure 10. It can be seen that the average performance gain increases initially and levels off with higher aggregated interpolation ratio. It reveals that the optimal aggregated interpolation ratio is ir1+ ir2+ ir4+ ir8. Further, this also confirms that using higher than optimal aggregated interpolation ratio does not necessarily improve the segmentation performance.

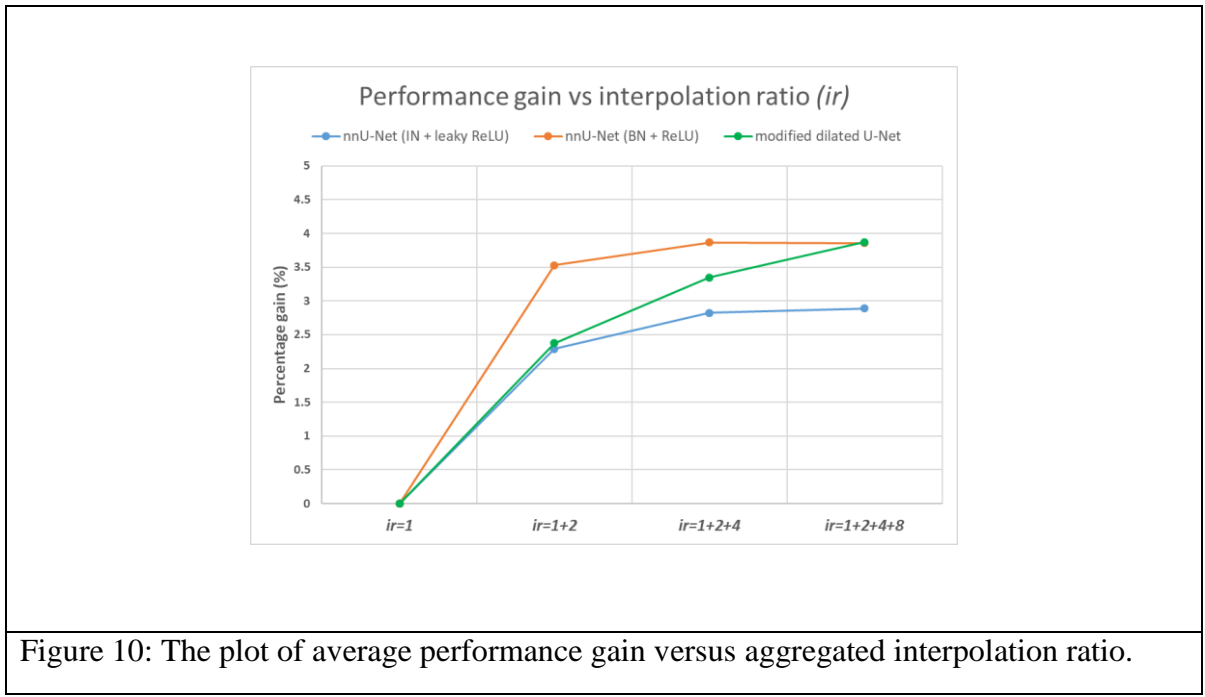

Figure 10: The plot of average performance gain versus aggregated interpolation ratio.

*Effect of ensemble learning strategy*

The effect of the ensemble learning strategy can be visualised by investigating 3D segmented airway masks. Figures 11 shows the selected 3D masks of airway segmentation for nnU-Net (IN + leaky ReLU) – case 3, nnU-Net (BN + ReLU) – case 7 and modified dilated U-Net – case 5. For case 3, the segmentation improvement can be observed from subsegmental bronchi to bronchioles. Regarding case 7, the segmentation of bronchi is gradually improved from ir1 to ir1 + ir2 + ir4 + ir8. The connection between lobar bronchi is also improved. Further, more higher generation bronchioles are segmented. The segmentation of trachea is improved for case 5. Additionally, some segmental bronchi are better segmented.

| | (A) nnU-Net (IN + leaky ReLU) Case 3 | (B) nnU-Net (BN + ReLU) Case 7 | (C) modified dilated U-Net Case 5 |
|---|---|---|---|
| Ground-truth | 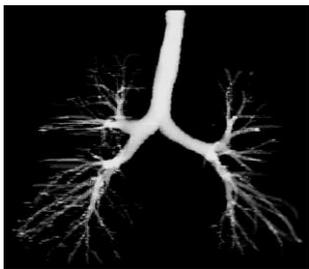 | 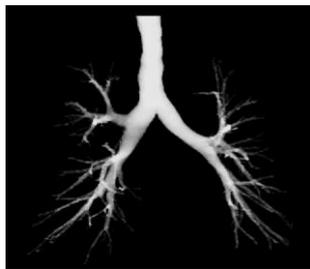 | 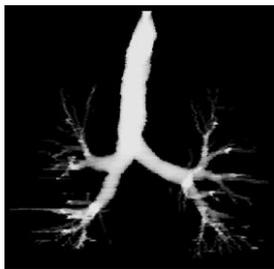 |
| Baseline | 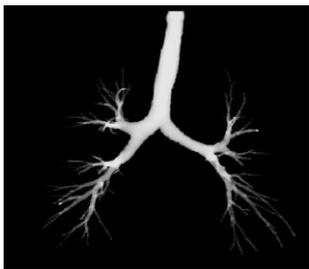 | 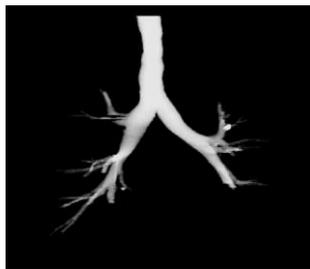 | 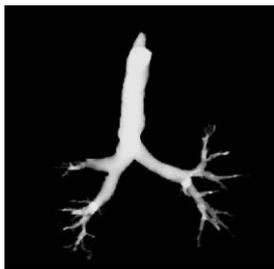 |
| ir1+ir2 | 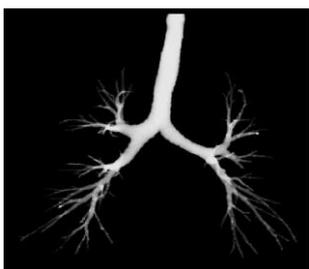 | 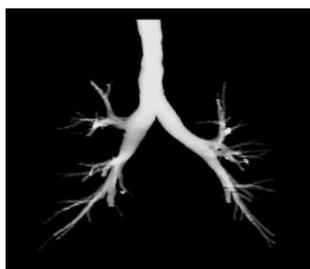 | 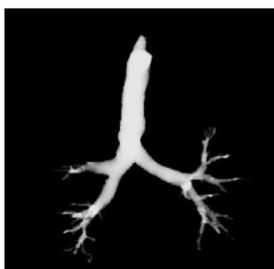 |
| ir1+ir2+ir4 | 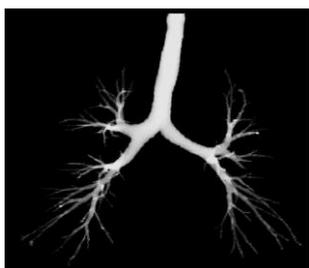 | 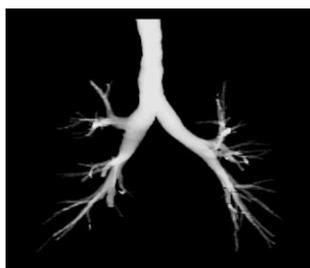 | 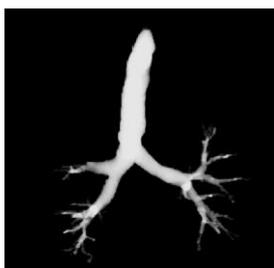 |
| ir1+ir2+ir4+ir8 | 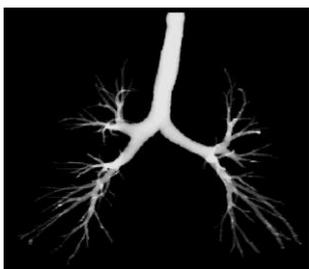 | 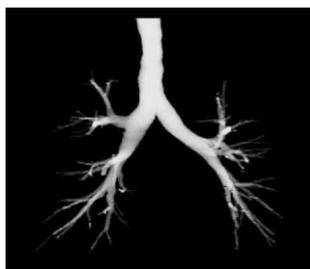 | 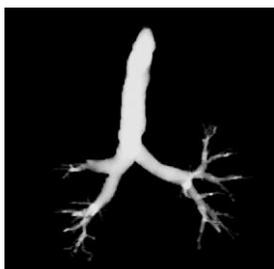 |

Figure 11: The selected 3D airway segmentation: (Column A) nnU-Net (IN + leaky ReLU) - Case 3, (Column B) nnU-Net (BN + ReLU) - Case 7, (Column C) modified dilated U-Net - Case 5.

*Effect of individual interpolation ratio (ir)*

The effect of individual interpolation ratio for cases 1, 4 and 5 is illustrated in Figure 12. By observing the segmented airways from ir1 to ir8, more bronchioles are segmented. Furthermore, when the highest interpolation ratio ($ir = 8$) is used, the segmentation of the trachea is the worst. In general, more artefacts are observed when a higher interpolation ratio is used.

|  | Case 1 | Case 4 | Case 5 |
|---|---|---|---|
| ir1 | 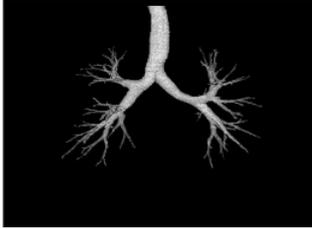 | 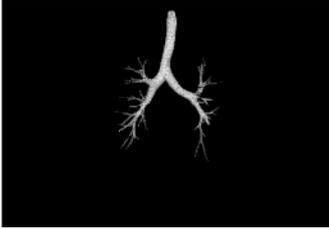 | 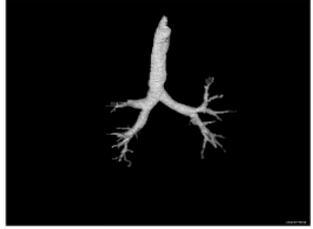 |
| ir2 | 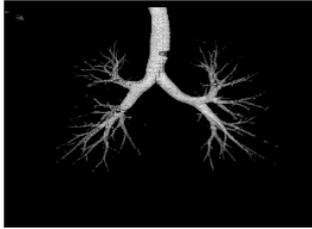 | 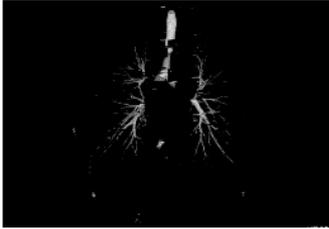 | 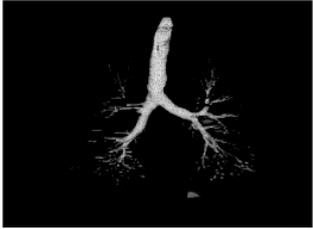 |
| ir4 | 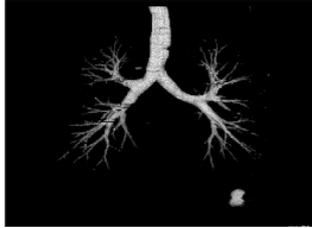 | 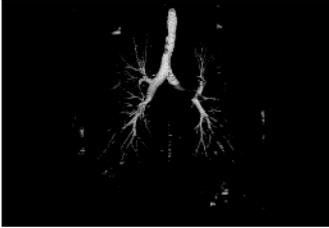 | 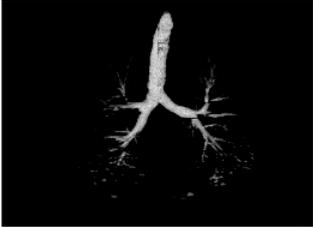 |
| ir8 | 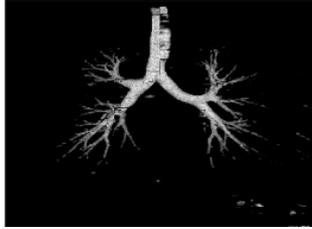 | 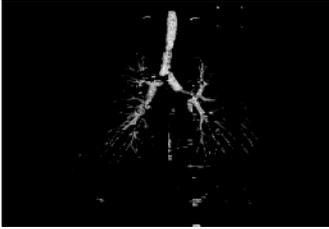 | 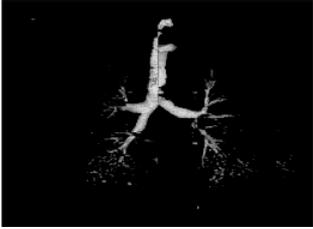 |

Figure 12: Effect of individual interpolation ratio for cases 1, 4 and 5.

*Blur effect*

The blur effect of our method is illustrated in Figure 13. The blur level is increased with increasing interpolation ratio. It is visually evident when the interpolation ratio is set at 4 and 8. Though the size of the bronchiole is increased after interpolation, the sharpness of the bronchiole wall is reduced. Further, the blur effect is not visually evident when the interpolation ratio is set at 2.

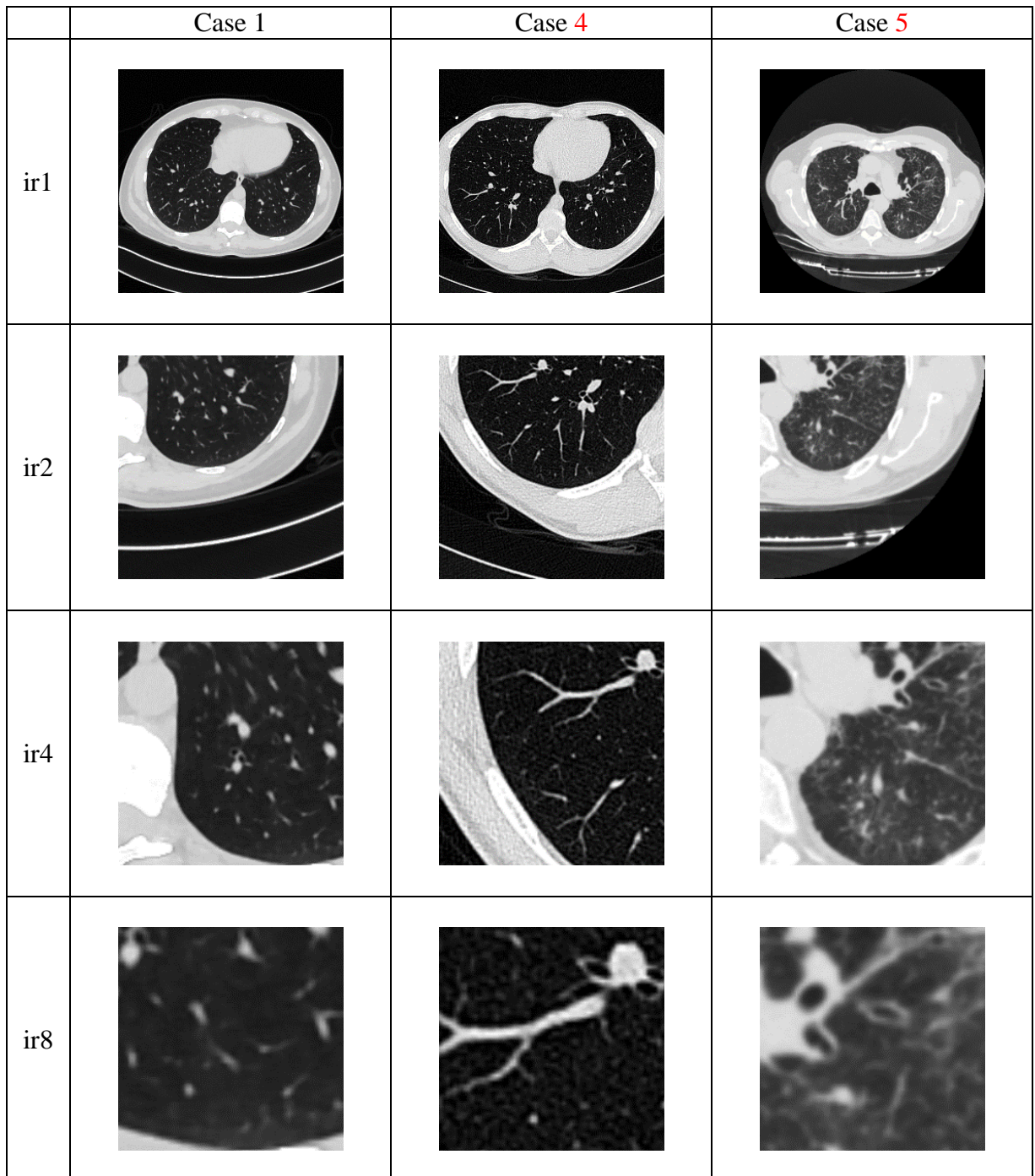

Figure 13: The blur effect of our method for cases 1, 4 and 5.

Regarding the blur effect of our method, a sharpening filter (Figure 14) can be used to reduce this effect and further improve the segmentation accuracy by about 1%.

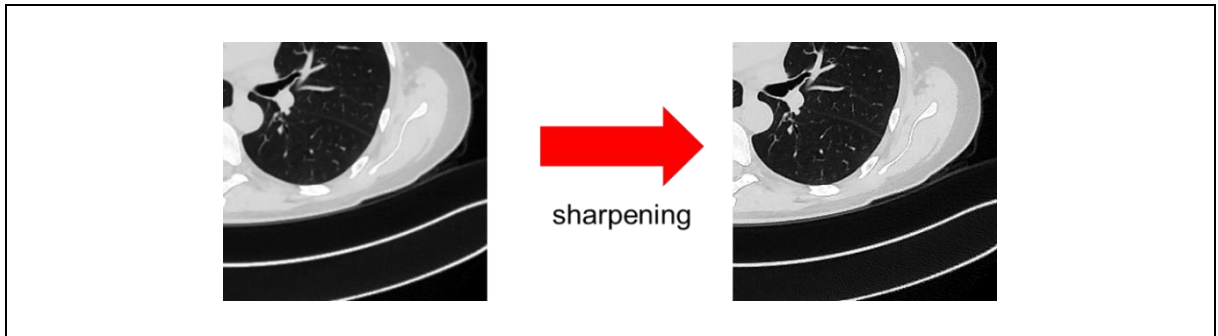

Figure 14: Implementation of sharpening filter on blurred image.

**Discussion**

A data-centric deep learning method with big interpolated data has been developed to improve airway segmentation on high resolution CT images. The proposed method can be applied to any 2D deep learning model including standard models such as the U-Net. Our study shows that the airway segmentation performance gain is between 0.21% and 14.11% using our Interpolation-Split.

The proposed method is good at improving (1) the connectivity between airway segments (2) airway wall segmentation (3) bronchi and bronchioles segmentation. It utilises zoom-in images and aggregates the segmented airways at different scales. The zoom-in images are useful for the model to capture the features of the walls of large airways and segment more small airways which are shape and scale/size dependent [33]. Furthermore, the ensemble learning strategy combines the airway segmentation at various interpolation ratios and hence improves the connectivity between airway segments.

In this study, we observe that the interpolation ratio affects the airway segmentation. Although more small airways are detected and segmented, the large airways such as the trachea and primary bronchi are not segmented well at higher interpolation ratios. This implies that an optimal scale/size range of airways exits for a given interpolation ratio. The higher interpolation ratio shifts the optimal scale/size range towards smaller airways.

It should be noted that the current study uses the threshold (0.5) for binarization. We also observe that changing the interpolation ratio affects the threshold. A further study is required to investigate the relationship between optimal threshold and interpolation ratio. We also noted that the sample size increases significantly with higher interpolation ratios and hence the training time increases accordingly. Data parallelism can be deployed to speed up the training and maintain the computational efficiency.

Our proposed technique requires low RAM (i.e., 16-64GB) usage when interpolation is performed. The GPU memory requirement is also low (i.e., 8/11 GB GDDR6) as the models have the low GPU memory utilization and the size of input image is fixed. Further, our Interpolation-Split is GPU memory efficient, because the GPU memory requirement does not increase throughout the pre-processing (including interpolation/split), training, validation and prediction stages. It only requires disk space to store the original/interpolated images, zip compression can be used to compress the images and save the disk space when the computational resources is low.

In this study, we use a 2D segmentation strategy for 3D CT volume and it is adopted from Zhang et al. [34]. Zhang et al. analysed a set of 2D MRI images extracted from 3D MRI volume. Then, these 2D images were fed into the 2D CNN deep learning model for multi-modality isointense infant brain image segmentation. Their approach outperformed existing methods and showed that deep learning model (2D CNNs) could produce more objective and accurate computational results for infant tissue image segmentation. Additionally, 2D CNN has lower computational cost compared with 3D CNN.

It should be noted that a small segmentation performance loss (-0.22%) was observed for case 1 when nnU-Net (IN + leaky ReLU) is used. This might be explained by the fact that ir1+ir2+ir4+ir8 is not the optimal configuration and lead to degraded segmentation. The optimal configuration for this case is ir1+ir2 and its segmentation performance gain was 0.12%. In general, ir1+ir2+ir4+ir8 is still the optimal configuration for all other cases.

A human tracheobronchial tree has 23 airway generations on average [35, 36]. High-resolution CT has the ability to image a smaller component of the airway tree as bronchioles with a diameter less than 2mm are not visible on HRCT. In healthy subjects up to 8 airway generations may be visible on HRCT [37], and the number of visible airway generations increases in disease states. The segmentation performance of healthy subjects was compared with IPF patients. Notably, our proposed method shows better performance gain on IPF patients. This might be explained by the observation that more abnormal small airways (between the 9th and 13th airway generations) [38] are found in IPF patients. This also reveals that our method improves the segmentation of small airways.

In this study, nnU-Net and modified dilated U-Net were chosen to be the baseline models. While our previous study [39] evaluated the segmentation performance on standard U-Net and its performance was about 75%. This also demonstrates the benefits and usefulness of the proposed technique applied to a more complex model.

Our study has several limitations. First, the subjects and patients were selected retrospectively. This might introduce the bias on data selection. Second, manual annotation was performed to produce ground-truth labels for airway tree segmentation. The annotators might bias the accuracy of the ground-truth labels. Third, the segmentation performance metric, DSC, might provide a biased measurement as the large and small airways were examined together. Larger airways segmented well might have resulted in a good DSC even if small airways were segmented poorly.

## Conclusion

Our study is the first to demonstrate the feasibility of using a data-centric deep learning method with big interpolated data to segment the airway tree resulting in a good segmentation performance gain. Our method requires low RAM/GPU memory usage; it also maintains GPU memory efficiency and has the flexibility to be deployed in any 2D deep learning model. Future work should investigate 3D data-centric approaches for their segmentation performance.

## Acknowledgement

JJ was supported by Wellcome Trust Clinical Research Career Development Fellowship 209553/Z/17/Z and the NIHR Biomedical Research Centre at University College London. This research was funded in whole or in part by the Wellcome Trust [209553/Z/17/Z]. For the purpose of open access, the author has applied a CC-BY public copyright licence to any author accepted manuscript version arising from this submission.

## Declaration of Interest

JJ declares fees from Boehringer Ingelheim, F. Hoffmann-La Roche, GlaxoSmithKline, NHSX, Takeda, Wellcome Trust, Microsoft Research unrelated to the submitted work and UK patent application numbers 2113765.8 and GB2211487.0.

## References

[1]   W. K. Cheung et al., "Automated airway quantification associates with mortality in idiopathic pulmonary fibrosis," *European Radiology,* vol. 33, no. 11, pp. 8228-8238, 2023/11/01 2023, doi: 10.1007/s00330-023-09914-4.
[2]   A. Pakzad et al., "Evaluation of automated airway morphological quantification for assessing fibrosing lung disease," *arXiv preprint arXiv:2111.10443,* 2021.
[3]   M. Zhang et al., "Multi-site, Multi-domain Airway Tree Modeling (ATM'22): A Public Benchmark for Pulmonary Airway Segmentation," *arXiv preprint arXiv:2303.05745,* 2023.
[4]   A. F. Frangi, W. J. Niessen, K. L. Vincken, and M. A. Viergever, "Multiscale vessel enhancement filtering," (in English), *Medical Image Computing and Computer-Assisted Intervention - Miccai'98,* vol. 1496, pp. 130-137, 1998, doi: DOI 10.1007/bfb0056195.
[5]   S. You, E. Bas, and D. Erdogmus, "Extraction of samples from airway and vessel trees in 3D lung CT based on a multi-scale principal curve tracing algorithm," *Annu Int Conf IEEE Eng Med Biol Soc,* vol. 2011, pp. 5157-60, 2011, doi: 10.1109/IEMBS.2011.6091277.
[6]   H. H. Duan, J. Gong, X. W. Sun, and S. D. Nie, "Region growing algorithm combined with morphology and skeleton analysis for segmenting airway tree in CT images," *J Xray Sci Technol,* vol. 28, no. 2, pp. 311-331, 2020, doi: 10.3233/XST-190627.

[7]   V. Badrinarayanan, A. Kendall, and R. Cipolla, "SegNet: A Deep Convolutional Encoder-Decoder Architecture for Image Segmentation," (in English), *Ieee T Pattern Anal,* vol. 39, no. 12, pp. 2481-2495, Dec 2017, doi: 10.1109/Tpami.2016.2644615.

[8]   K. Sun *et al.*, "High-resolution representations for labeling pixels and regions," *arXiv preprint arXiv:1904.04514,* 2019.

[9]   O. Ronneberger, P. Fischer, and T. Brox, "U-Net: Convolutional Networks for Biomedical Image Segmentation," (in English), *Medical Image Computing and Computer-Assisted Intervention, Pt Iii,* vol. 9351, pp. 234-241, 2015, doi: 10.1007/978-3-319-24574-4_28.

[10]  F. Milletari, N. Navab, and S. A. Ahmadi, "V-Net: Fully Convolutional Neural Networks for Volumetric Medical Image Segmentation," (in English), *Int Conf 3d Vision,* pp. 565-571, 2016, doi: 10.1109/3dv.2016.79.

[11]  C. Shorten and T. M. Khoshgoftaar, "A survey on Image Data Augmentation for Deep Learning," (in English), *J Big Data-Ger,* vol. 6, no. 1, Jul 6 2019, doi: 10.1186/s40537-019-0197-0.

[12]  S. Budd, E. C. Robinson, and B. Kainz, "A survey on active learning and human-in-the-loop deep learning for medical image analysis," *Med Image Anal,* vol. 71, p. 102062, Jul 2021, doi: 10.1016/j.media.2021.102062.

[13]  J. P. Charbonnier, E. M. van Rikxoort, A. A. A. Setio, C. M. Schaefer-Prokop, B. van Ginneken, and F. Ciompi, "Improving airway segmentation in computed tomography using leak detection with convolutional networks," (in English), *Med Image Anal,* vol. 36, pp. 52-60, Feb 2017, doi: 10.1016/j.media.2016.11.001.

[14]  J. Yun *et al.*, "Improvement of fully automated airway segmentation on volumetric computed tomographic images using a 2.5 dimensional convolutional neural net," (in English), *Med Image Anal,* vol. 51, pp. 13-20, Jan 2019, doi: 10.1016/j.media.2018.10.006.

[15]  S. A. Nadeem, E. A. Hoffman, and P. K. Saha, "A Fully Automated CT-Based Airway Segmentation Algorithm using Deep Learning and Topological Leakage Detection and Branch Augmentation Approaches," (in English), *Proc Spie,* vol. 10949, 2019, doi: 10.1117/12.2512286.

[16]  Y. L. Qin, Y. Gu, H. Zheng, M. J. Chen, J. Yang, and Y. M. Zhu, "Airwaynet-Se: A Simple-yet-Effective Approach to Improve Airway Segmentation Using Context Scale Fusion," (in English), *I S Biomed Imaging,* pp. 809-813, 2020. [Online]. Available: <Go to ISI>://WOS:000578080300161.

[17]  K. Zhou *et al.*, "Automatic airway tree segmentation based on multi-scale context information," (in English), *Int J Comput Ass Rad,* vol. 16, no. 2, pp. 219-230, Feb 2021, doi: 10.1007/s11548-020-02293-x.

[18]  A. Garcia-Uceda, R. Selvan, Z. Saghir, H. A. W. M. Tiddens, and M. de Bruijne, "Automatic airway segmentation from computed tomography using robust and efficient 3-D convolutional neural networks," (in English), *Sci Rep-Uk,* vol. 11, no. 1, Aug 6 2021, doi: 10.1038/s41598-021-95364-1.

[19]  H. Zheng *et al.*, "Alleviating Class-Wise Gradient Imbalance for Pulmonary Airway Segmentation," (in English), *Ieee T Med Imaging,* vol. 40, no. 9, pp. 2452-2462, Sep 2021, doi: 10.1109/Tmi.2021.3078828.

[20]  J. Q. Guo *et al.*, "Coarse-to-fine airway segmentation using multi information fusion network and CNN-based region growing," (in English), *Comput Meth Prog Bio,* vol. 215, Mar 2022, doi: 10.1016/j.cmpb.2021.106610.

[21]  C. L. Wang *et al.*, "Tubular Structure Segmentation Using Spatial Fully Connected Network with Radial Distance Loss for 3D Medical Images," (in English), *Medical Image Computing and Computer Assisted Intervention - Miccai 2019, Pt Vi,* vol. 11769, pp. 348-356, 2019, doi: 10.1007/978-3-030-32226-7_39.

[22]  A. G. U. Juarez, R. Selvan, Z. Saghir, and M. de Bruijne, "A Joint 3D UNet-Graph Neural Network-Based Method for Airway Segmentation from Chest CTs," (in English), *Machine*

*Learning in Medical Imaging (Mlmi 2019),* vol. 11861, pp. 583-591, 2019, doi: 10.1007/978-3-030-32692-0_67.

[23] Y. Q. Wu, M. H. Zhang, W. H. Yu, H. Zheng, J. S. Xu, and Y. Gu, "LTSP: long-term slice propagation for accurate airway segmentation," (in English), *Int J Comput Ass Rad,* vol. 17, no. 5, pp. 857-865, May 2022, doi: 10.1007/s11548-022-02582-7.

[24] S. Chen *et al.*, "Label Refinement Network from Synthetic Error Augmentation for Medical Image Segmentation," *arXiv preprint arXiv:2209.06353,* 2022.

[25] M. Zhao *et al.*, "GDDS: Pulmonary Bronchioles Segmentation with Group Deep Dense Supervision," *arXiv preprint arXiv:2303.09212,* 2023.

[26] J. Parker, R. V. Kenyon, and D. E. Troxel, "Comparison of interpolating methods for image resampling," *IEEE Trans Med Imaging,* vol. 2, no. 1, pp. 31-9, 1983, doi: 10.1109/TMI.1983.4307610.

[27] A. Mumuni and F. Mumuni, "Data augmentation: A comprehensive survey of modern approaches," *Array,* vol. 16, p. 100258, 2022/12/01/ 2022, doi: https://doi.org/10.1016/j.array.2022.100258.

[28] S. Yang, W. Xiao, M. Zhang, S. Guo, J. Zhao, and F. Shen, "Image data augmentation for deep learning: A survey," *arXiv preprint arXiv:2204.08610,* 2022.

[29] P. Lo *et al.*, "Extraction of airways from CT (EXACT'09)," *IEEE Trans Med Imaging,* vol. 31, no. 11, pp. 2093-107, Nov 2012, doi: 10.1109/TMI.2012.2209674.

[30] A. Thelen, S. Frey, S. Hirsch, and P. Hering, "Improvements in shape-from-focus for holographic reconstructions with regard to focus operators, neighborhood-size, and height value interpolation," *IEEE Trans Image Process,* vol. 18, no. 1, pp. 151-7, Jan 2009, doi: 10.1109/TIP.2008.2007049.

[31] F. Isensee, P. F. Jaeger, S. A. A. Kohl, J. Petersen, and K. H. Maier-Hein, "nnU-Net: a self-configuring method for deep learning-based biomedical image segmentation," (in English), *Nat Methods,* vol. 18, no. 2, pp. 203-+, Feb 2021, doi: 10.1038/s41592-020-01008-z.

[32] F. Yu and V. Koltun, "Multi-Scale Context Aggregation by Dilated Convolutions," *CoRR,* vol. abs/1511.07122, 2016.

[33] W. K. Cheung, "State-of-the-art deep learning method and its explainability for computerized tomography image segmentation," *Explainable AI in healthcare: Unboxing machine learning for biomedicine*, M. S. Raval, M. Roy, T. Kaya, and R. Kapdi, Eds.: Chapman and Hall/CRC, 2023. [Online]. Available: https://doi.org/10.1201/9781003333425-5

[34] W. Zhang *et al.*, "Deep convolutional neural networks for multi-modality isointense infant brain image segmentation," *NeuroImage,* vol. 108, pp. 214-224, 2015/03/01/ 2015, doi: https://doi.org/10.1016/j.neuroimage.2014.12.061.

[35] A. Bouhuys, *The Physiology of Breathing: A Textbook for Medical Students*. Grune & Stratton, 1977.

[36] E. R. Weibel, *Morphometry of the Human Lung*. Springer, 1963.

[37] A. A. Diaz *et al.*, "Airway Count and Emphysema Assessed by Chest CT Imaging Predicts Clinical Outcome in Smokers," *Chest,* vol. 138, no. 4, pp. 880-887, 2010/10/01/ 2010, doi: https://doi.org/10.1378/chest.10-0542.

[38] S. E. Verleden *et al.*, "Small airways pathology in idiopathic pulmonary fibrosis: a retrospective cohort study," *Lancet Respir Med,* vol. 8, no. 6, pp. 573-584, Jun 2020, doi: 10.1016/S2213-2600(19)30356-X.

[39] M. Abbas, "Automatic Segmentation of Bronchiectasis Affected Lungs Using UNETs on Lung Computed Tomography Imaging," *Thesis, MEng in Computer Science, UCL Computer Science, University College London,* 2020.